\documentstyle[a4,11pt,epsf]{article}
\begin{document}
\begin{Large}
\begin{center}
{\bf The Extremely High Energy Cosmic Rays}
\end{center}
\end{Large}

\bigskip
\bigskip

\centerline{{\sf Shigeru Yoshida$^a$ and Hongyue Dai$^{b,}$\footnote{
Present Address: Rosetta Inpharmatics,12040 115th Ave NE,
Kirkland, WA 98034,USA} }}

\bigskip
\centerline{$^a$Air Shower Division, Institute for Cosmic Ray Research,}
\centerline{University of Tokyo, 3-2-1 Midori-cho, Tanashi, Tokyo 188, Japan} 

\centerline{$^b$ High Energy Astrophysics Institute, Department of Physics}
\centerline{ University of Utah, Salt Lake City, UT 84112,USA}

\begin{abstract}

Experimental results from Haverah Park, Yakutsk, AGASA and Fly's Eye
are reviewed. All these experiments work in the energy range above
$10^{17}$ eV.  The 'dip' structure around $10^{18.5}$ eV in the energy
spectrum is well established by all the experiments, though the exact
position differs slightly.  Fly's Eye and Yakutsk results on the
chemical composition indicate that the cosmic rays are getting lighter
over the energy range from $10^{17}$ eV to $10^{19}$ eV, but the exact
fraction is hadronic interaction model dependent, as indicated by the
AGASA analysis. The arrival directions of cosmic rays are largely
isotropic, but interesting features may be starting to emerge.  Most of
the experimental results can best be explained with the scenario that
an extragalactic component gradually takes over a galactic population
as energy increases and cosmic rays at the highest energies are
dominated by particles coming from extragalactic space.  However,
identification of the extragalactic sources has not yet been
successful because of limited statistics and the resolution of the
data.

\end{abstract}

\vspace{1cm}

{\large subject headings: cosmic rays: general}

\vspace{5mm}

\parindent=15.0pt
\section{Introduction}

Many kinds of radiation exist in the universe, including photons
and particles with a wide range of energies. Some of the
radiation is produced in stars and galaxies, while some is
cosmological background radiation, a relic from the history of
cosmic evolution.  Among all this radiation, the most energetic
are cosmic ray particles: nucleons, nuclei, and even extremely
energetic gamma rays. Their energies appear to reach beyond
$10^{20}$ eV. Cosmic rays with energies above $10^{19}$ eV were
first detected by the Volcano Ranch group led by John Linsley of
the University of New Mexico more than 30 years ago.  Since then,
where and how these particles are produced and how they propagate
in space have been puzzles. Their extremely high energies, seven
orders of magnitude greater than those of any nucleons that
humans have thus far been able to accelerate on earth, 
suggest that unbelievably
energetic phenomena have occurred somewhere in the universe. One
problem is that the extremely low flux at these energies (the
typical rate of cosmic rays above $10^{20}$ eV is one
event/km$^2$/century!)  requires a detector with a huge acceptance
which has always been challenging to build due to
technological and economical difficulties.  The pioneer detector
at Volcano Ranch covered an area of only 9 $km^2$.
Experiments following Volcano Ranch have improved in
terms both of statistics and data quality.  Throughout these
years of continuous effort, signatures concerning the origin of
these Extremely High Energy Cosmic Rays (EHECRs) have started to
emerge.

In this paper we review the current situation of our understanding of
the origin of EHECRs, based mainly on recent observational results.  
To help our interpretation of the data, we first discuss the conditions
required for a site to be a source of EHECRs, and briefly describe how
these cosmic rays propagate through space.  Next the techniques of
detecting these particles are summarized and the major detectors
are introduced. Following that, the recent experimental results
are reviewed in terms of the energy spectrum, chemical composition, and
anisotropy.  A two component model, which we think is highly
reasonable, is constantly checked against these results. Finally, the
consistency of the results compared with the simple model
is summarized.

\subsection{Extremely High Energy Cosmic Rays: General View}

\begin{figure}
\centerline{{\epsfxsize9cm\epsfbox{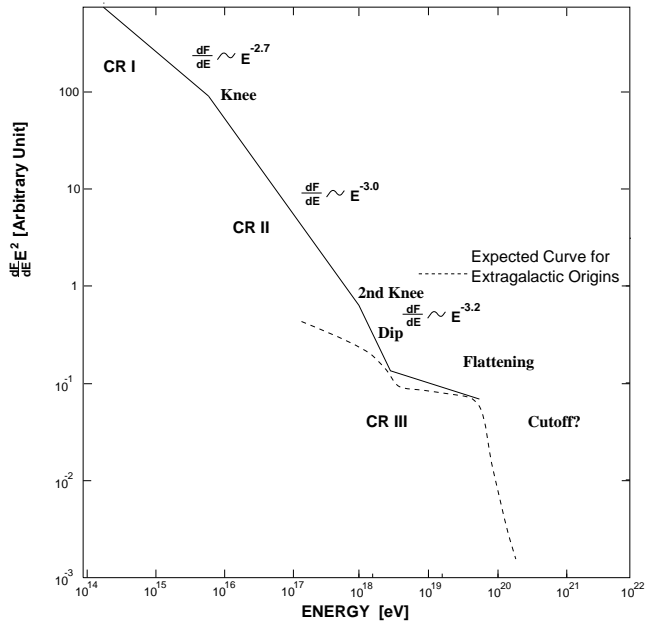}}}
\caption{A schematic drawing of the energy spectrum of cosmic rays
above $10^{14}$ eV.}
\label{fig:schematic_spec}
\end{figure}

The energy spectrum of high energy cosmic rays above 10 GeV
(where the magnetic field of the sun is no longer a concern) is
well represented by a power law form. Figure
\ref{fig:schematic_spec} is a schematic drawing of the energy
spectrum. In terms of its structure, the spectrum can be divided
into three regions: two ``knees'' and one ``ankle''.  The first
``knee'' appears around $3\times10^{15}$ eV where the spectral
power law index changes from -2.7 to -3.0. The second ``knee'' is
somewhere between $10^{17}$ eV and $10^{18}$ eV where the
spectral slope steepens from -3.0 to around -3.3.  The ``ankle''
is seen in the region of  $3\times10^{18}$ eV. Above that energy, 
the spectral slope flattens out to about -2.7, however,
there is large uncertainty due to
poor statistics and resolution.  Our interest in this
paper is this final and most energetic population, the
EHECRs. The production of the first two populations is likely to
be explained with conventional first order Fermi shock
acceleration \cite{drury91} at energetic objects such as
supernova remnants within our galaxy, although many concerns about
the effectiveness remain.  The third population is interesting
since it raises the following difficult questions: How do they get such
huge energies?  Where do they come from?  Does the spectrum end
somewhere?  What is the chemical composition of these cosmic
rays?

\subsection{The Possible Sites to Produce EHECRs}

\begin{figure}
\centerline{{\epsfxsize7cm\epsfbox{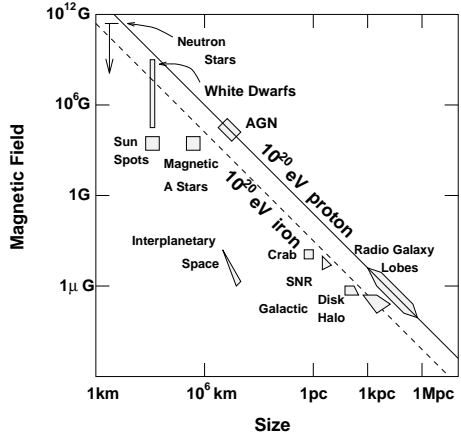}}}
\caption{The Hillas Diagram showing possible sites
for acceleration of cosmic rays up to $10^{20}$ eV.}
\label{fig:accel_sites}
\end{figure}

The existence of cosmic rays with energies 
up to nearly $10^{20}$eV has been solidly
established, but the acceleration theories are on much less
solid footing.
No matter how the particles are accelerated, the upper
bound of the energy gained should be determined by balancing
the acceleration time with escape time from the acceleration
site. In 1984, Hillas proposed the following constraint
\cite{hillas84}:

$$ B_{\mu G}L_{pc} >{E\over 10^{15}\ eV}{1\over Z\beta}, \eqno(1)$$

where $L_{pc}$ is the size of the site in parsec, $Z$ is charge
of the particle, and $\beta$ is the speed of the scattering
waves within the field of the site.  The magnetic field needs to
be large enough to confine the particles within their
acceleration site and the size of the site must be sufficiently
large for particles to gain sufficient energy before they
escape. These simple requirements already rule out most
astronomical objects in the universe.  Figure
\ref{fig:accel_sites} shows objects which satisfy this
dimensional requirement.  Most of the galactic objects are
excluded simply because they are too small and/or have magnetic fields
that are too weak.  Only a few extragalactic objects such as
Active Galactic Nuclei (AGN) and radio galaxies remain as
possible candidates.  This fact is the basic reason many favor the
extragalactic origin of EHECRs.

In any actual acceleration site, energy loss mechanisms
always compete with the gain of energy.  With first order Fermi
shock acceleration, the acceleration time is proportional to the
mean free path for scattering in the shock wave, which itself is
approximately inversely proportional to the magnetic field
strength. Therefore, a certain magnitude of magnetic field is
required, not only to confine the particles within the site, but
also to accelerate the particles quickly.  However, too strong a
magnetic field also causes problems for particle acceleration,
because it can cause protons to lose energy via synchrotron
radiation.  Other strong energy losses are caused by collisions
with photons and/or matter at the acceleration site. A certain
photon field is normally expected at the site as a result of
synchrotron radiation by electrons or thermal radiation off the
accretion disk.  This leads to the additional requirement,
that the site must have sufficiently low densities of 
radiation and matter so that
cosmic ray nucleons are able to accelerate 
to $\sim 10^{20}$ eV before
losing significant energy. 
This raises more difficulties with candidate sites. 
For example, the core region of AGNs are ruled out because for this
reason. The relativistic jets found in some classes of
AGNs such as blazars may be able to produce EHECRs
\cite{ip91,mannheim95}, although models require optimistic
fine-tuning of the acceleration efficiency and the Doppler boost
factor of the relativistic jets.  Rachen and Biermann have
proposed hot spots of Fanaroff-Riley type II galaxies as EHECR
sources \cite{rachen93}.  There seems to be enough acceleration
power with not too-dense photons at the hot spots.  However, the
possibility is not excluded that collisions with UV photons in
the spots discourage the acceleration of protons above $10^{20}$
eV.  While relativistic jets of blazars
and hot spots of FR II type radio galaxies are candidates sources of
EHECRs via a conventional first order Fermi shock
acceleration mechanism, it is not obvious that the acceleration
efficiency is large enough to produce particles up to $10^{20}$ eV.

These extragalactic models favor protons for the EHECR
composition. Heavier nuclei like iron may break up into
nucleons by photodisintegration during the shock acceleration,
through collisions with UV photons at the site.

\bigskip

The difficulties in acceleration can be avoided if EHECRs are
direct products of processes which do not require acceleration.
``Top down'' scenarios have recently been proposed
\cite{bhattacharjee92,bhat95} involving relics of
symmetry-breaking phase transitions in the early universe such as
cosmic strings and magnetic monopoles, so called topological defects.
If such defects exist, they may have produced EHE particles with
energies up to the grand unified theory (GUT) scale (typically
$\sim 10^{25}$ eV) through the decay of the X-particles released
in the collapse or decay of the defects.  Because the hadron jets
created at the decay of the X-particle are the main channels of
particle production in this model, neutrinos and gamma rays,
rather than protons and neutrons, are predominant.  Any heavy
nuclei like iron are completely ruled out in this model because
the hadron jets create no nuclei.  Propagation effects in the
cosmic background radiation field (described later) would modify
the emitting spectrum of each component, but one would still
expect that gamma rays may be dominant at energies above $10^{20}$ eV,
with details dependent on the strengths of the universal radio
background and the extragalactic magnetic field, both of which
are poorly known.  The basic problem in this scenario, is that
topological defects are exotic: the absolute intensity of 
defects is unknown.  The observed intensities of cosmic rays and
diffuse gamma rays can put constraints on the upper bound of
defect intensity.  So far, we have no experimental evidence, however,
measurement of an excess of $\gamma$ ray flux above $10^{20}$ eV
and detection of EHE neutrinos above $10^{19}$ eV would be
signatures of topological defects.

It has been suggested that Gamma Ray Bursts (GRB), responsible
for gamma rays up to the GeV range, may also be able to produce
EHECRs.  This would be a burst source and 
not a continuously emitting one.  This would also result in
a correlation between arrival times and energies of EHECRs.
Unfortunately, the time scale might be much longer than any single
experiment can afford to run and thus the correlation may be
extremely difficult to detect. Detail on this idea is presented
in \cite{waxman95,vietri95}.

\subsection{Propagation of EHECRs in space}

It is important to understand how EHECRs propagate from their
sources to earth, since this puts constraints on possible sources
and provides hints for the most effective way of searching for
them.  First, the galactic magnetic field of $\sim\mu$G can no
longer confine cosmic ray protons with energies greater than
$10^{19}$ eV in the galactic disk 
since the Larmor radius of a proton at that energy,
$$ L_{kpc} \simeq 
({E\over 10^{19}\ eV})({B\over 3\ \mu G})^{-1} \times 3 kpc,
\eqno(2) $$
becomes greater than the thickness of our galactic
disk.  This means that any galactic protons can easily escape
from our galaxy, provided that the galactic magnetic fields 
do not extend out into a possible Halo.
This again favors the hypothesis of an energetic
extragalactic component dominating galactic components in
the EHECRs population.

Secondly, when EHECRs are traveling through extragalactic space,
their trajectories are not strongly bent by the extragalactic
magnetic field and the arrival directions of such cosmic rays
should point back to their emitters.  Information on the
extragalactic magnetic field strength is difficult to gather.
We know only the Faraday rotation bound on the extragalactic
magnetic field is given by \cite{kronberg94}
$$ B_{rms}\sqrt{l_c} \leq 10^{-9} G\ Mpc^{1\over 2} \eqno(3) $$
where $l_c$ is the scale of the coherent magnetic field in Mpc,
and the mean deflection angle can be written as
$$ \theta_{def} \leq ({R\over 10Mpc})^{1\over 2} 
({E\over 3\times 10^{19}\ eV})^{-1} \times 3.2^{\circ} \eqno (4) $$
for protons \cite{waxman96}. Here $R$ is distance to the source.
This opens a new window of astronomy, that of Charged Particle
Astronomy.

A typical deflection angle of $\sim 3^{\circ}$, which is
comparable with the typical angular resolution of the present
experiments, might be still too large in an actual search for
sources, because there exist many astronomical objects 
in a $3^{\circ}\times 3^{\circ}$ window even if the candidates are
limited to AGNs and radio galaxies. The real situation is
much better, however, since there is a limit on the distance
over which an EHECR may travel.
We can limit the search to relatively nearby sources,
because EHECRs collide with
cosmological backgrounds and lose energy during their
propagation.  This is the most important effect on the
propagation of EHECRs.

\begin{figure}
\centerline{{\epsfxsize7cm\epsfbox{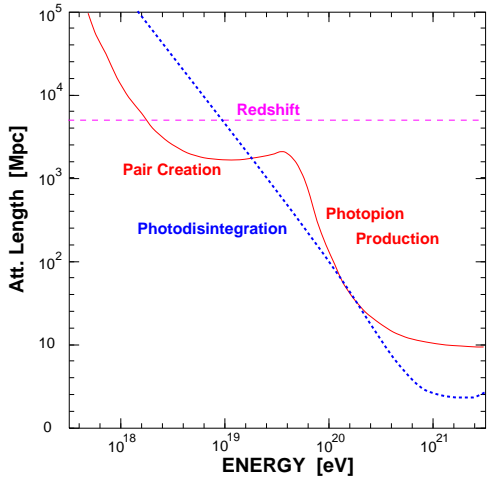}}}
\caption{The attenuation length of cosmic rays as a function of energy.
The solid curve shows the case for nucleons calculated by Yoshida and Teshima.
The dashed curve shows the case for iron calculated 
by Puget, Stecker and Bredekamp. The bound given by redshift 
(adiabatic energy loss) are applicable to all primaries.}
\label{fig:att_length}
\end{figure}

\begin{figure}
\centerline{{\epsfxsize6cm\epsfbox{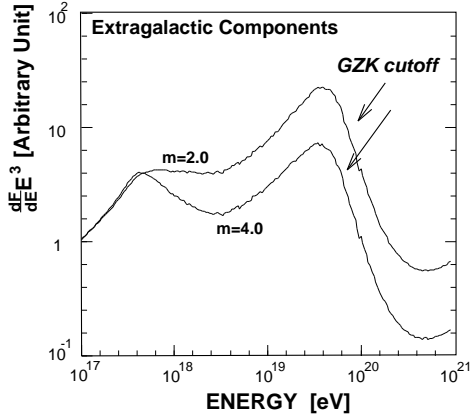}}}
\caption{The expected spectral shapes for two assumptions
concerning the distribution of sources in extragalactic space.
$m$ is the evolution parameter of EHECR emission. A larger $m$
means more contributions from sources in distant (high redshift)
regions.  This extragalactic population might dominate in the
highest energy region of the observed spectrum above $\sim
10^{19}$ eV. The galactic magnetic field, which could suppress
the the recovery below $10^{18.5}$ eV, is not included in the
simulation.}
\label{fig:spec_proton_dif}
\end{figure}

\bigskip
EHECR protons or neutrons interact with the microwave
background photons through pair creation and photopion
production.  The threshold energy of photopion production, the main
energy loss process, is
$$ E \simeq 7\times 10^{19} ({E_{bb}\over 2\times 10^{-3}\ eV})^{-1}
(1+\cos\theta)^{-1}\quad eV \eqno(5) $$
where $E_{bb}$ is the energy of the microwave background photon,
from a blackbody spectrum with a characteristic temperature of
$2.7^{\circ}$ K.  The photon and the EHECR interact with a
collision angle $\theta$.  Above the threshold energy, EHECRs
rapidly lose energy. This may 
result in a cutoff in the energy spectrum. This
cutoff is known as Greisen-Zatsepin-Kuzmin cutoff (GZK cutoff)
\cite{grei66,zats66}, and is the centerpiece of the EHECRs
physics.  A detection of this effect proves the extragalactic
origin of EHECRs and limits the distance to possible sources to
less than $\simeq 100$ Mpc for particles above $10^{20}$ eV.  The
situation is explained in figure \ref{fig:att_length} which shows
the attenuation length of protons in extragalactic space.  One
finds that the attenuation length above the photopion production
threshold is contracted by rapid energy losses. The attenuation
length for protons with energies higher than $7\times 10^{19}$ eV
(the threshold energy of photopion production) is shorter than
500 Mpc \cite{yosh93,prot96}.  Any sources contributing to the
bulk of EHECRs above this energy should be within 500 Mpc of
earth.  The higher the energy, the shorter the upper bound on the
distance.  A $3\times 10^{20}$ eV proton would require sources
within only $\sim 50$ Mpc. Thus,
nearby sources should make the dominant contributions 
at the high energy end of the spectrum.
This effect provides an important feature on the resulting energy
spectrum {\it shape}.  Because the microwave background during
cosmological evolution is a function of redshift, the spectral
shape of EHECRs also depends on the redshift, as well as the
source distribution in space.  Figure \ref{fig:spec_proton_dif}
shows the expected spectral shapes if many sources are
isotropically distributed in the universe \cite{yosh93}.  The
parameter $m$ describes the cosmological evolution of cosmic ray
emission.  Therefore, it controls the relative contributions of
sources at different distances.  The spectral shape changes with
the parameter $m$, however, the cutoff energy remains near $5\times
10^{19}$ eV. The dominant contribution of nearby sources at the
high energy end make the spectral shape above $10^{19}$ eV less
sensitive to cosmological effects.  The shape around the GZK
cutoff is universal while most of the cosmological signatures are
found in the $10^{17}-10^{18}$ eV region where another cosmic ray
population may dominate.  A search for a cutoff at around
$5\times 10^{19}$ eV is indeed a robust method for obtaining
evidence of the extragalactic origin regardless of details in the
model. 

A similar situation exists for primary nuclei, like carbon or
iron.  This time, photodisintegration is the limiting factor
rather than photopion production. As a result,
there is an even more rapid energy
loss during propagation as shown in figure \ref{fig:att_length}
\cite{puge76}.  
Since heavy nuclei break down quickly during propagation,
an EHECR composition favoring protons and neutrons is likely.
EHE nuclei will be reduced both at the acceleration sites and over
the propagation volume.

\begin{figure}
\centerline{{\epsfxsize9cm\epsfbox{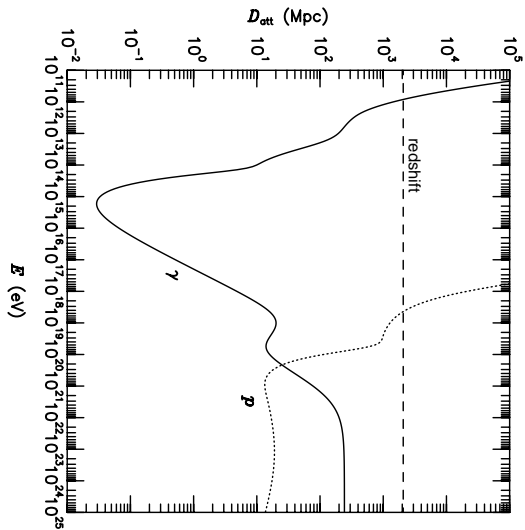}}}
\caption{The attenuation length of photons traveling in
extragalactic space.}
\label{fig:photon_att}
\end{figure}

\bigskip
We should also consider the possibility of EHE cosmic rays being
photons.  The most conventional mechanism for the production of
EHE photons is the decay of neutral pions produced by a collision
between an EHE cosmic ray nucleon and a background photon during
propagation.  The exotic ``top down'' scenario involving
topological defects predicts a more predominant initial photon
flux \cite{sigl96}.  These EHE photons/electrons initiate
electro-magnetic cascades on a low energy radiation field such as
the microwave background.  The attenuation length of photons in
the radiation field of extragalactic space is shown in figure
\ref{fig:photon_att} \cite{lee96}.  EHE photons interact with
microwave/radio background photons via pair creation and double
pair creation. Electrons produced in this process transfer most
of their energy to a background photon via inverse Compton
scattering or sometimes via triplet pair production ($e\gamma_b
\to e e^+e^-$).  Since the EHE $\gamma$ ray attenuation length
does not decrease with energy (as is the case for protons), there
is no cutoff feature in the spectrum \cite{sigl96,lee96}.  This
leads to the prediction of a dominant gamma ray flux at energies
above the GZK cutoff.  It should be pointed out that the $\gamma$
ray flux depends on two poorly known parameters: the
extragalactic magnetic field and the universal radio background.
A strong radio field reduces the mean free path for pair
creation, and synchrotron radiation cools the electron pairs
out of the EHE range.  The conventional shock
acceleration models always predict very low gamma ray fluxes
\cite{yosh93} while the ``top down'' models provide a
possibility of $\gamma$ ray dominance around $10^{20}$
eV\cite{sigl96}.

\begin{figure}
\centerline{{\epsfxsize7cm\epsfbox{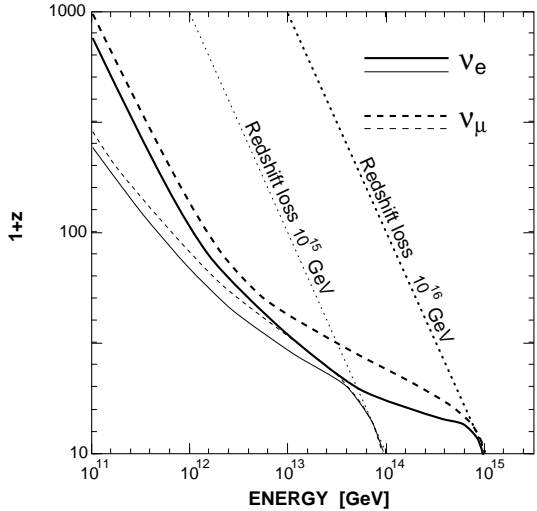}}}
\caption{The horizon of the universe for EHE neutrinos as a
function of present-day neutrino energies. Two cases are shown
for neutrino primary energies at emission of $10^{15}$ GeV (thin
lines) and $10^{16}$ GeV (thick lines), both of which could be
reasonable in the top down model of EHECR production involving
topological defects.}
\label{fig:neut_horizon}
\end{figure}

\bigskip
Finally, we discuss the case of neutrinos.  EHE neutrinos can
certainly be created through the decay of charged pions produced
by collisions between EHE cosmic ray nucleons and microwave
background photons \cite{yosh93,hill83,yoshida97}.  These
secondary neutrinos are good probes of EHE particle emission
activities at early epochs of the universe, since their flux
strongly depends on the evolution parameters.  The detection of
these neutrinos is unfortunately a remote possibility since the
most optimistic flux is only comparable to that of observed EHE
cosmic rays.  Nevertheless, a search for neutrinos in the EHE
range (above $\sim 10^{19}$ eV) is a meaningful test of the
topological defect hypothesis since that model 
predicts a much higher neutrino flux \cite{yoshida97,sigl97}.  
Because the maximum energy of neutrinos reaches 
the GUT scale ($\sim 10^{25}$ eV) and
their emission at superhigh redshift epochs ($z\sim 500$) are the
main contributions in this scenario, collisions of EHE neutrinos
with low energy cosmological relic neutrinos are not
negligible \cite{roulet93,yoshida94}.  Figure
\ref{fig:neut_horizon} shows the horizon of the universe for EHE
neutrinos, the maximum redshift to which EHE neutrinos are not
attenuated in their propagation \cite{yoshida94}.  The dotted
lines correspond to the upperbound of the horizon when one
considers the redshift energy loss only.  It is shown that
interactions with the relic neutrinos, which contract the
horizon, are a key effect in the ``top down'' scenario because
the emissions of EHE neutrinos at high redshift epochs ($z>100$)
are predominant due to the higher rate of annihilation of
topological defects.  It should be noted that some EHE neutrinos
that initiate neutrino cascading on the cosmological neutrino
background field will further enhance 
the EHE neutrino flux at earth, and
the planned future experiments may be able to detect a few of
them \cite{yoshida97}.  These interactions also
play an important role in the recently proposed
mechanism to generate observable particles above the GZK cutoff
\cite{weiler97} by collisions of EHE cosmic ray neutrinos with
possibly clustered massive neutrinos in our galaxy.

\section{Experiments}

   The fundamental questions in cosmic ray physics are the origin
and production mechanisms of the cosmic rays. To answer these
questions, every experiment in the extremely high energy region,
almost without exception, measures three quantities - the primary
energy, composition and arrival direction. We will survey the
measurements of these quantities.

   This report focuses on cosmic rays with energies greater than
10$^{17}$ eV.  At such high energies all the measurements are
indirect due to the extremely low flux.
A high energy primary particle enters the atmosphere and
interacts with air molecules initiating a cascading process that
produces secondary particles. The result is called an air
shower and only the secondary
particles from these air showers are actually detected.
When it reaches the ground, the footprint of an air
shower can cover an area of tens of square kilometers. The
secondary particles also collide with and 
excite nitrogen molecules in the
air, and thereby provide a flash of fluorescence light (the
light being emitted by the de-excitation of the nitrogen
molecules) in the atmosphere. 

There are two main types of detectors: the ground array and
the air fluorescence detector.  The ground array experiments
sample the charged secondary particles as they reach the ground,
and determine the primary energy from the particle density, the
arrival direction from detector trigger times, and may 
infer the primary chemical composition from the
ratio of the muon to electron component.  Another type of
detector, the air fluorescence detector, views tracks of light in
the atmosphere. It determines the track geometry either by the
photomultiplier tube trigger times or by so-called ``stereo''
reconstruction, then calculates the primary energy by an integral
along the track length, and deduces the chemical composition by
the shape of the longitudinal shower development. There is
copious Cherenkov light produced along the shower axis by the
charged particles, and a large area Cherenkov array can be used to
detect that light.  The total flux of Cherenkov light is a good
measure of the total particle track integral in space and is thus a
good primary energy parameter. The angular and lateral distribution of
Cherenkov light can be used to deduce the primary composition.

   Serious research on extremely high energy cosmic rays started
with the Volcano Ranch experiment \cite{lin63a,lin63b} more than
30 years ago, subsequently joined by the Sydney SUGAR array
\cite{brow68,winn86}, Haverah Park \cite{wats91,lawr90,lawr91},
Yakutsk \cite{efim91}, Fly's Eye \cite{bal85a,bal85b}, AGASA
\cite{naga92,yoshida95} and HiRes \cite{abu97a,abu97b,abu97c}
experiments.  The currently running experiments are Yakutsk,
AGASA and HiRes, with new experiments at the proposal stage
including Auger \cite{cron92,auge97,bora97,daw97a}, the Telescope
Array \cite{ais97a,ais97b}, OWL \cite{orme97} and Space Air Watch
\cite{lins97}.
In this report, we will discuss the Fly's Eye, AGASA, Haverah
Park and Yakutsk experiments.

\subsection{The Haverah Park Experiment}

    The Haverah Park Experiment was an array of water Cherenkov tanks
situated at 53$^o$58.2' N, 1$^o$38.2' W, at an altitude of 200 m above
sea level. It was operated from 1968 to 1987 by the University of
Leeds and other U.K. groups. The water tanks were 1.2 m deep, with
stations ranging in area from 1 to 54 m$^2$ enclosing a ground array
of approximately 12 km$^2$. The array trigger requirement was a signal
equivalent to that produced by ten vertical muons in the central 34
m$^2$ detector coincident with 1 similar signal in one of the three
other 34 m$^2$ detectors 500 m from the center. This produced an array
threshold of approximately 6$\times$10$^{16}$eV for vertical showers.

    The Haverah Park experiment developed the technique of
using the particle density
600 m from the shower core ($\rho(600)$, \cite{hill71}) to determine
the primary energy. For a given primary energy, the particle density
at large distances is believed to have smaller fluctuations than
densities nearer the shower core, the latter being related to
fluctuations in shower development. The parameter $\rho$(600) is also
insensitive to the primary composition and, to some extent,
insensitive to the interaction model used to derive the relation
between the primary energy and $\rho(600)$. It is a more robust energy
parameter then the total ground shower size.

\subsection{The Yakutsk Experiment}

    The Yakutsk array is situated in Russia at longitude 129.4$^o$E
and latitude 61.7$^o$ N. It was expanded in 1973 to increase 
the sensitivity to
EHE cosmic rays. It covers a ground area of approximately 20
km$^2$. Over the period of operation, new detectors have been added to
make the array denser.  In 1973, there were 44 plastic scintillation
detectors, 35 of which were 4$m^2$ in area, with 9 at 2$m^2$ area. 
By 1992, there
were 58 plastic scintillation detectors in total. Among these
detectors, 49 are 4$m^2$ and 9 are 2$m^2$\cite{afan96}.  Detectors are
arranged on a triangular grid. The spacing of the outer detectors is
approximately 1km, with the center of the array filled with detectors
on a 500 m triangular grid. Two trigger schemes exist. One is formed
by the outer detectors and is sensitive to showers with energies above
10$^{18}$ eV (trigger-1000). The second trigger 
is formed by the detectors on the
500m grid and has a threshold energy of approximately 10$^{17}$ eV
(trigger-500).

  Underground muon detectors have been gradually added to the
array. There are now 5 underground detectors, each of 20 $m^2$ area,
with a muon threshold energy of $E_{\mu} \ge 1\times sec\theta$ GeV,
and one underground detector of 192 $m^2$ area with $E_{\mu} \ge
0.5\times sec\theta$ GeV ($\theta$ being the zenith angle of the
particle).

   The array is also equipped with 50 Cherenkov detectors for studies of both
pulse widths and the total integrated light. This latter quantity is used as a
check of the method of converting the ground parameter S600 (the particle
density 600 m from the shower core, measured by the scintillation detectors) to
primary energy. The S600 resolution is estimated to be $\le$ 20\%
\cite{egor93,afan96} for vertical showers. 
Since the primary energy is linearly
proportional to S600, the energy resolution should be of the same
order.

\subsection{The AGASA experiment}

The Akeno Giant Air Shower Array (AGASA) is located at the Akeno
observatory in Japan ($35.8^{\circ}$N $138.5^{\circ}$E).  The array
consists of 111 scintillators of 2.2 $m^2$ area located on the surface to
measure the charged particle densities and 27 sets of proportional
counters under absorbers to measure the muon component of air
showers. The threshold energy of the muon detectors is approximately
0.5 GeV. The AGASA array covers an area of 100 $km^2$ and is the
world's largest detector now in operation. At least 95 \% of the
detectors have been operated since 1991 and full operation began in
1993.  All the detectors are connected to an optical fiber network so
that their operation, monitoring, calibration and triggering can be
controlled remotely \cite{yoshida95}.  Recently a new data acquisition
system was installed to unify all the triggering over the entire
array \cite{ohoka97}. As a result, 
the detection aperture for air showers with
energies greater than $10^{19}$ eV became $\sim 125 km^2 sr$ (events
with zenith angles less than $45^{\circ}$), about 60 \% larger than
before the new system was installed.

The method of energy assignment is based on S600. Experimental
uncertainties in the measurement of this energy estimator have been
carefully studied using their own data \cite{yoshida94b} and the energy
resolution is estimated to be $^{+18}_{-25}$ \% on average for events
with E$\geq 10^{19}$ eV. The largest uncertainty arises from poor
understanding of the attenuation of S600 as a function of zenith
angle at higher energies.  However, the measurement of the attenuation
is now being improved with the increase in event numbers.  This will
lead to a better estimate of resolution in the near future.

\subsection{The Fly's Eye Experiment}

The Fly's Eye detectors\cite{bal85a,cass85} were located at Dugway,
Utah ($40^o$N, $113^o$W, atmospheric depth 860 g $cm^{-2}$). The
original detector, Fly's Eye I, consisted of 67 spherical mirrors of
1.5 m diameter, each with 12 or 14 photomultipliers at the focus. The
mirrors were arranged so that the entire night sky was imaged, with
each phototube viewing a hexagonal region of the sky 5.5 degrees in
diameter. Fly's Eye I began full operation in 1981. In 1986 a second
detector (Fly's Eye II)  was completed 3.4 km away.
Fly's Eye II consisted of 36 mirrors of the same design.  
This detector only viewed
the half of the night sky in the direction of Fly's Eye I. Fly's Eye
II could operate as a stand alone device or in conjunction with Fly's
Eye I for a stereo view of a subset of the air showers. There were 880
photomultiplier tubes in Fly's Eye I and 464 tubes in Fly's Eye II.

The Fly's Eye tubes detected nitrogen fluorescence light, and direct
and scattered (by Rayleigh and Mie scattering) \v{C}erenkov light.  Of
these, fluorescence relates most directly to the local number of
charged particles in the air shower.  The nitrogen fluorescence light
is produced in the spectral region 310 to 440 nm and is emitted
isotropically from the shower, allowing for detection of showers at
large distances.

The Fly's Eye detector was the first successful air fluorescence
shower detector and showed its power in energy resolution and in
composition resolution. Because the detector viewed the shower
development curves, its energy estimation is almost totally
interaction model independent. The development curves are also 
able to put
constraints on hadronic interaction models used in composition
studies.

An extremely important feature of the Fly's Eye detector was that
those showers viewed by both Fly's Eye I and II (i.e. ``stereo'' events)
were measured with significant redundancy.  This provided a model
independent way of checking the energy and 
depth of shower maximum ($X_{max}$) resolution. 
The energy and $X_{max}$ values were independently
 reconstructed from each eye using
the stereo geometry (assuming the stereo geometry is well determined).
Be comparing the results,
the stereo energy resolution was determined to be 24\% for events
below $2\times10^{18}$ eV and 20\% for events above $2\times10^{18}$
eV. The $X_{max}$ resolution for the stereo events is 47 $g/cm^2$. The
monocular energy resolution was calculated by comparing the monocular
energy with the stereo energy event by event, and the FWHM is
estimated to be 36\% for events below $2\times10^{18}$ eV and 27\% for
events above $2\times10^{18}$eV. It should be pointed out that the
monocular energy resolution is underestimated using this method since
the stereo energy reconstruction also uses the Fly's Eye I phototube
intensities.  In the stereo case, the energy resolution ($(E_1 -
E_2)/E_{avg}$) follows a Gaussian distribution, but in the monocular
case, only $log(E_{mono}/E_{stereo})$ is Gaussian.  Here $E_1$ and
$E_2$ are the energies determined from Fly's Eye I and II
independently using the stereo geometry. $E_{avg}$ is the average of
$E_1$ and $E_2$.  $E_{mono}$ is the energy determined by the monocular
reconstruction using Fly's Eye I information only, and $E_{stereo}$ is
the energy determined by stereo reconstruction using information from
both eyes. Although the monocular and stereo energy resolution figures
look quite similar in terms of FWHM, the monocular resolution function
has a much longer tail. Its effect on the spectrum is significant, as
we will see in the next section.

\section{The structure of the energy spectrum }

Because of the steeply falling spectrum and the limited energy
resolution, the features of the cosmic ray energy spectrum are usually
described with pictorial but not very scientific names like ``knee'' and
``ankle''.

It has been known for a long time that apart from the ``knee'' (where
the spectral index changes from -2.7 to -3.0 at around
$3\times10^{15}$eV) , the spectrum changes shape and flattens again
around 10$^{19}$ eV ( the so called ``ankle''). Like the ``knee'', the
exact shape of the ``ankle'' is very uncertain. There is another feature
which is not usually mentioned - between the ``knee'' and the ``ankle'',
there is another change of slope, around a few times $10^{17}$ eV.

\begin{figure}
\centerline{{\epsfxsize8.0cm\epsfbox{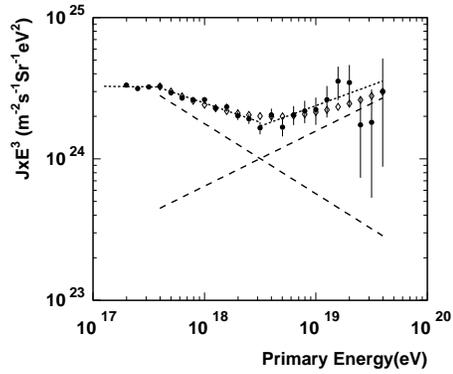}}}
\caption{ Fly's Eye stereo energy spectrum. Dots: data. Dotted line: best
fit in each region. Dashed line: a two component fit.}
\label{fig:fly_stereo}
\end{figure}

In our opinion the best results on the ``ankle'' structure come from the
Fly's Eye stereo data (Fig.\ref{fig:fly_stereo}) \cite{bird93,bird94},
because of the well controlled error estimates of that data set. The
spectrum becomes steeper immediately beyond $10^{17.6}$ eV and
flattens beyond $10^{18.5}$ eV. The change in the spectral slope forms
a dip centered at $10^{18.5}$ eV. The slopes for each segment and for
the whole spectrum are listed in Table 1.  To show the significance of
the dip, the Fly's Eye group calculated the expected number of events
based on overall single slope spectrum, but normalized to the
intensity at $10^{17.6}$ eV.  The expected number of
events between $10^{17.6}eV$ and $10^{19.6}eV$ was 5936.3, compared
with the observed number, 5477. The significance of this deficit is
6.0 $\sigma$.  To show the significance of the flattening above
$10^{18.5} eV$, the group used the normalization and slope from a
total fit up to $10^{18.5} eV $.  The total number of observed events
above this energy is 281 while the expected number would be 230, a
3.4$\sigma$ excess.  The excess is even more pronounced (5.2$\sigma$)
if the spectrum from $10^{17.6}$ to $10^{18.5}$eV is used to calculate
the expectation (in this case, the expectation is 205.9 events above
$10^{18.5}$eV).  As noted above, the energy resolution 
is estimated to be 24\% for
events below $2\times10^{18}$ eV and 20\% for events above.  The
flattening is therefore not the result of a resolution effect (an
improving energy resolution would make the spectrum steeper).
The existence of the dip is further supported by the Fly's Eye raw
event energy distribution \cite{bird94}. The fact that this
distribution also shows a dip excludes the possibility that the dip is
artificially introduced by the aperture calculation.

\begin{table}[b]
\begin{center}
\caption{ Normalizations and spectral slopes of J(E)}
\begin{tabular}{cccc}\hline
Energy range (eV) & Power index & 
$\log (normalization [m^{-2}\sec^{-1}sr^{-1}eV^{-1}])$ & Normalized at\\
$10^{17.3} - 10^{19.6}$ & $-3.18\pm 0.01$ & -29.593 & $10^{18} eV$ \\
$10^{17.3} - 10^{17.6}$ & $-3.01\pm 0.06$ & -29.495 & $10^{18} eV$ \\
$10^{17.6} - 10^{18.5}$ & $-3.27\pm 0.02$ & -29.605 & $10^{18} eV$ \\
$10^{18.5} - 10^{19.6}$ &  $-2.71\pm 0.10$ &-32.623 & $10^{19} eV$ \\
\hline
\end{tabular}
\end{center}
\label{table:fly_spectral}
\end{table}

Although the clearest, the Fly's Eye experiment is not the only observation
to see the 'dip'. AGASA, Haverah Park and Yakutsk have reported
similar observations.

\begin{figure}
\centerline{{\epsfxsize7cm\epsfbox{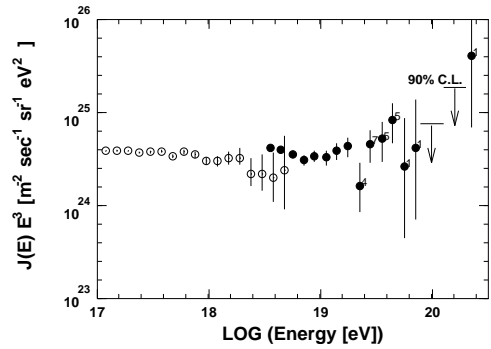}}}
\caption{ The energy spectrum measured by the Akeno 1km$^2$ array
(open circles) and AGASA (filled circles). The AGASA measurement
plotted here is their first published spectrum in 1995.}
\label{fig:energy_spec_all}
\end{figure}

The AGASA group has used a more densely packed array, known as the
Akeno $1km^2$ array (A1), to study cosmic rays below $3\times 10^{18}$
eV.  They have reported that the energy spectrum starts quite steeply
at $10^{17.8}$ eV with spectral index of $3.2\pm 0.1$
\cite{naga92}. This behavior was confirmed by the first published
measurement by AGASA in 1995 \cite{yoshida95}. Figure
\ref{fig:energy_spec_all} shows the measured spectrum. It is seen that
the AGASA spectrum also shows the flattening around $10^{18.8}$ eV. A
likelihood analysis calculated a spectral index above $10^{18.8}$ eV
of $2.7^{+0.2}_{-0.4}$.  The significance of this flattening is
2.9$\sigma$. More recent data from AGASA has increased this
significance to 3.2$\sigma$ \cite{takeda97}. 

\begin{figure}
\centerline{{\epsfxsize8.0cm\epsfbox{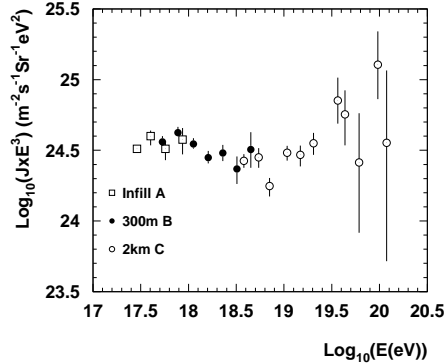}}}
\caption{The Haverah Park energy spectrum.}
\label{fig:haverah_e}
\end{figure}

     Haverah Park reported the steepening of the spectrum above
$5\times10^{17}$ eV as early as 1980 \cite{cunn80}. Their final
energy spectrum \cite{wats91,lawr91} (Fig.\ref{fig:haverah_e})
confirms that result with better statistics.  Using a subset of
the data (set B in \cite{lawr91}), they measured the differential
energy spectrum slope between $10^{17.7}$ to $10^{18.7}$ eV and
found it to be $3.24\pm0.07$, very similar to the Fly's Eye
stereo slope. By including 8 more good quality events above
$10^{18.7}$ eV, their maximum likelihood analysis gave a
differential slope of $3.14^{+0.05}_{-0.06}$.  Given this slope the
group expect 65.5 events above $10^{19}$ eV and actually observe
106 events.  The significance of the spectral flattening is
therefore 5$\sigma$.

The Haverah Park final spectrum consists of three different data sets
with different selection criteria, and hence different energy
resolutions. Their set B, which was used to derive the slopes, has a
very strict cut on the event geometry and therefore has the best
control over errors.  The estimated resolution of $\rho$(600) for this
data set is better than 15\%. For the spectrum above $4\times10^{18}$
eV all data available are included because of the extremely low flux.
However each event was manually checked to make sure that the routine
fits are reasonable.

\begin{figure}
\centerline{{\epsfxsize8.0cm\epsfbox{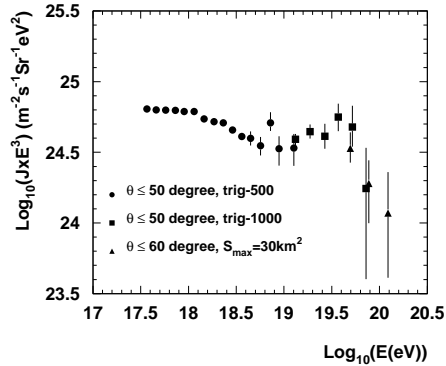}}}
\caption{The Yakutsk energy spectrum.}
\label{fig:yakutsk_e}
\end{figure}

The recent Yakutsk
spectrum\cite{afan95a,afan96}(Fig.\ref{fig:yakutsk_e}) confirms the
general shape reported by the Fly's Eye stereo spectrum, except that
the dip position is moved up by approximately 0.3 in logarithmic terms
(roughly at $10^{18.85}$eV) near where AGASA saw the dip. 
The significance
of the dip depends on which spectral slope is assumed, but within the
range of fitting errors, the group showed that the deficit between
$10^{18.1}$ and $10^{19.6}$ eV is 7.9 $\sigma$ and the excess above
$10^{18.85}$ eV (``ankle'') is estimated to be 1.9 $\sigma$ or 6.6
$\sigma$ depending on how the spectrum is extrapolated.

Among all of these experiments, Fly's Eye gave the lowest spectral
normalization, and Yakutsk gave the highest (they differ by nearly a
factor of 2.5 around $10^{18}$ eV). The difference could be due to
three potential problems: the absolute energy calibration, the exposure
calculation, or the energy resolution. It is obvious how the energy
calibration and aperture estimation affect the flux. The effect of the
energy resolution is discussed by Sokolsky, Sommers and
Dawson\cite{soko92}.

Yakutsk has proposed to use the ``dip'' position to cross-calibrate
the energy among the experiments\cite{afan96}. We believe this is not
a good idea, since energy resolution effects can shift the position of
the ``dip''. The ``dip'' position can be pushed up in energy by the
``downhill'' fluctuation of lower energy events.

\section{ Possible explanations of the structure}

    The Fly's Eye group has proposed a simple, two component,
model to explain the structure of the spectrum
\cite{bird93}. The cosmic ray spectrum above $10^{17}$ eV is
considered as a superposition of two components, a steeply
falling component ( spectral index of -3.5) and a flatter
component (spectral index -2.6). The group further assumed that
the steep falling component originated in our own Galaxy and the
flatter component, which is dominant above a few times $10^{18}$
eV, has an extragalactic origin.

This two component hypothesis can be supported by examining the
whole energy spectrum above $10^{13}$ eV.  Examining
Fig. \ref{fig:schematic_spec}, we see that below the flatter
spectral component which appears at around a few times $10^{18}$
eV, all the other spectral features are steepenings.
Inefficiencies in the acceleration and/or the confinement in our
galaxy could cause all of these steepenings without the necessity
for new component.  However, the {\it flattening} is likely to
require the emergence of a new population.  The Fly's Eye group's
picture of a two component model is the most straightforward
interpretation of the energy spectrum without any complicated
model-dependent arguments.  Supernova remnants may
be suitable sites for the production of cosmic rays up to
energies at least to the ``knee''. Therefore the low energy component could
be associated with galactic origin while the extremely high
energies of the flat component would suggest that extragalactic
cosmic ray emitters are responsible for their production, as we
saw in the introduction.  The next question one should ask is
what other features are expected from the two component scenario.

First, the composition.  As we discussed in the introduction, the
Larmor radius of a $10^{18}$ eV proton is of the order of 300
parsec, comparable to the thickness of the galactic disk. That
means that the galactic disk cannot confine protons beyond this
energy.  Therefore, if the first component is of galactic origin,
it would need to be heavy.  From direct measurements we know that
the composition below the first ``knee'' is a mixture of light
and heavy primary particles. The composition should gradually get
heavier above the ``knee''  region due to likely
inefficiencies in acceleration or confinement.  What about the
second component?  Photodisintegration will essentially break up
any nuclei at the acceleration site, or over the course of
propagation if their energies are beyond $10^{20}$ eV. 
Thus, only nucleons and photons will be likely to survive.
It should be remembered that
exotic sources like topological defects will not produce heavy
nuclei either - only nucleons and photons.
 
What does the two component model say about the end of the
spectrum?  In this picture, the second component is of
extragalactic origin, and the particles are very likely to be
nucleons.  Therefore, we should see a GZK cutoff.  The spectrum
could recover after the cutoff, but a suppression in the flux
between the cutoff and the recovery should definitely be present.
The possible dominance of gamma rays as predicted in the
topological defects scenario might appear, but {\it above} the
cutoff.

Next, the anisotropy. Compared with protons, the Larmor radius
for heavy nuclei is reduced by a factor of the charge number Z,
which leads to a smaller anisotropy at energies leading up to
$10^{18}$ eV if the first component is dominated by heavy nuclei.
There may be a slight chance to see an anisotropy associated with
the galactic plane between $10^{18}$eV and $10^{19}$ eV before
the second component becomes dominant.  Anisotropies associated
with the second component would depend on the source distribution
and the magnetic field strength in extragalactic space. An
isotropic source distribution will most likely lead to an
isotropic arrival distribution.  However when the energy is high
enough, as we have discussed in the introduction, the GZK
mechanism will limit the source distances and the extragalactic
magnetic field will be incapable of bending the particle
trajectories too much.  Consequently we may be able to see some
anisotropy there.

Now let us check the consistency of these features with
observation.

\section{The spectrum near the Greisen-Zatsepin-Kuzmin cut-off}

  The GZK cutoff is an unambiguous feature expected in the spectrum
of extragalactic EHECRs. The search for this cutoff has been a major
aim for all of the experiments. 
The result of these studies is not yet clear.
From the point of view of the measurements, a cutoff feature
would appear in the form of a ``deficit'' in the number of events
above the expected cutoff energy.  This raises two major
difficulties in the search for the GZK cutoff.  These are
fundamental reasons for the lack of success so far in obtaining
concrete evidence for the existence of the cutoff.

\begin{figure}
\centerline{{\epsfxsize8cm\epsfbox{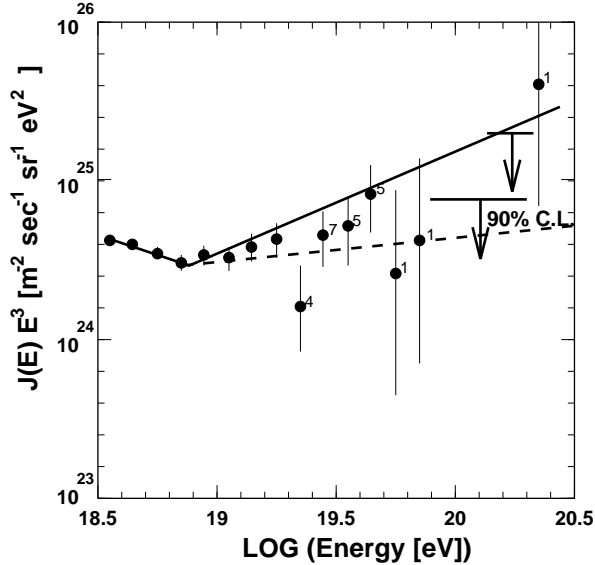}}}
\caption{Search for a cut-off in the measured spectrum.
The data points with error bars are measurements 
from the AGASA experiment published in 1995. }
\label{fig:cutoff_search}
\end{figure}

\begin{enumerate}

\item {\sf What is the definition of the ``deficit''?}

With limited statistics it is not easy to tell whether a deficit
is actually detected.  If this deficit were detected somewhere
near the middle of the energy range observed by a detector, then
an observation of a recovery from the deficit would be a
straightforward way to prove the existence of the deficit.
Unfortunately however, the GZK cutoff is expected at the end of
the spectrum where the statistics are extremely poor and where
the event ``deficit'' would produce even poorer statistics. The
traditional way to deal with this problem is to estimate the
expected number of events above a given energy, say $10^{20}$ eV,
and compare that number with number of observed events. However,
the {\it expected} number of events cannot be given {\em a
priori} by a source model but must be estimated from the spectrum
measurement itself, a process which contains many uncertainties.

Let us illustrate this situation in figure
\ref{fig:cutoff_search}.  The expected number of events depends
on how the spectrum is extrapolated from the lower energy
region. The extrapolation shown by the thick solid line gives
a deficit with greater than 90 \% confidence.  But
if one chooses another extrapolation, as shown by the dashed line
(which is also consistent at some level within statistical
errors), the conclusion is that this spectrum does not contain
any evidence of an event deficit.  Another concern is how one
determines the threshold energy for this test.  The number of
events above $10^{19.9}$ eV would favor the existence of a cutoff
in the example shown in figure\ref{fig:cutoff_search}, but if the
number of events above $10^{20.1}$ eV was examined instead there
would be no signature of any cutoff feature. As mentioned
previously, the model prediction for the GZK cutoff energy is
around $4\sim 5\times 10^{19}$ eV, but the actual energy where
the event deficit might appear will be higher than the prediction
because of limited energy resolution and possible
systematics. Therefore, the integration of the number of events above
$4\times 10^{19}$ eV, which can be justified by the theory, would
usually lower any statistical significance.  On the other
hand, any {\it optimization} of the threshold energy to give the
best significance has no justification at all.

\item {\sf Contamination of events with over-estimated energies}

The study of the GZK cutoff is very sensitive to the energy
resolution of each event.  The spectrum is so steep that even if
a small fraction of events suffer from an overestimation of
energy, there could be a significant effect on the spectral
shape. The tail of the energy error distribution could easily
smear out structures in the spectrum. This is especially the case
in the search for the cutoff, as one needs to search for an event
deficit which could easily disappear by contamination by only a
few events whose energies are overestimated. Any test of the GZK
cutoff hypothesis should pay great attention to limitations and
systematics in the energy resolution of the experiments.

\end{enumerate}

\begin{figure}
\centerline{{\epsfxsize8cm\epsfbox{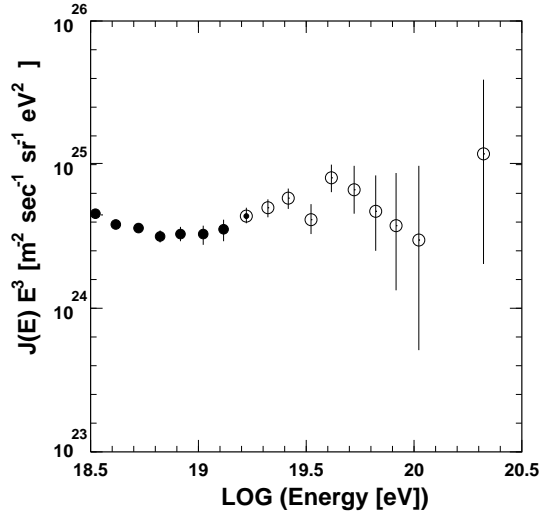}}}
\caption{The energy spectrum measured by the AGASA experiment.}
\label{fig:en_spec_agasa}
\end{figure}

The updated AGASA spectrum is shown in figure
\ref{fig:en_spec_agasa} \cite{hayashida97}.  To show the
significance of the GZK cutoff hypothesis in this spectrum, the
AGASA group analyzed their data in the following way.

\bigskip
If the expected spectrum curve with the GZK cutoff can be
approximately written as
$$ {dF\over dE}= \kappa E^{- \gamma}[1+\alpha f_c(E)], \eqno(1) $$
where $\kappa$ is a normalization factor,
$\gamma$ is the power law index, $f_c(E)$ is a function
to express the GZK cutoff term, and $\alpha$ is the contribution
coefficient of the cutoff term ($\alpha=0$ means no cutoff),
then a Likelihood function can be defined for the energy distribution
of the cosmic rays above 10 EeV in $\gamma - \alpha$
space as follows:

$$ L = \prod\limits_{i}^{all\ event} L_i(\gamma, \alpha) = 
\prod\limits_{i}^{all\ event} 
\int\limits_{10EeV} dE {dF\over dE}(\gamma, \alpha) 
{d\rho_i\over dE} A(E).
\eqno(2) $$
Here $A$ is the acceptance of the array to detect air showers, a
flat function above $10^{19}$ eV. The variable $\rho_i$ 
is the probability of
the primary energy of event $i$ being $E$, a function derived by
the AGASA event generators and analysis procedures.  The
parameter set ($\gamma_0,\alpha_0$) needed to maximize $L$ gives
the most likely value of $\gamma$ and $\alpha$. The confidence
level can be calculated with integration of the likelihood over
$\gamma$ and $\alpha$ space.

The formula for $f_c(E)$ comes from a modification of the
energy loss equation for photopion production calculated
by Berezinsky and Grigor'eva \cite{berezinsky88} as follows:
$$ f_c(E) = \exp[-\epsilon_c/E](\epsilon_c/E-1), \eqno(3) $$
where $\epsilon_c$ is the GZK cutoff energy, calculated to be
$4\times 10^{19}$ eV \cite{yosh93}.  The spectrum curve
calculated by equations (1) and (3), and that calculated by a
detailed numerical calculation \cite{yosh93} for the case of
isotropically distributed extragalactic sources, agree well
for the test of the GZK cutoff hypothesis.

This method is a fair approach for dealing with the difficulties
mentioned above. Using the spectrum power index $\gamma$ as a
free parameter automatically takes into account the uncertainties
in the spectrum shape.  Events do not have equal weight in the
analysis.  An event has a weight determined by an estimate of its
energy uncertainty. Contamination of the sample by a number
of poorly fitted events would not significantly change the
results. No uncountable degree of freedom,
such as energy bin width or the
choice of the threshold energy in the analysis, 
exists in the estimation of the significance. 

\begin{figure}
\centerline{{\epsfxsize8cm\epsfbox{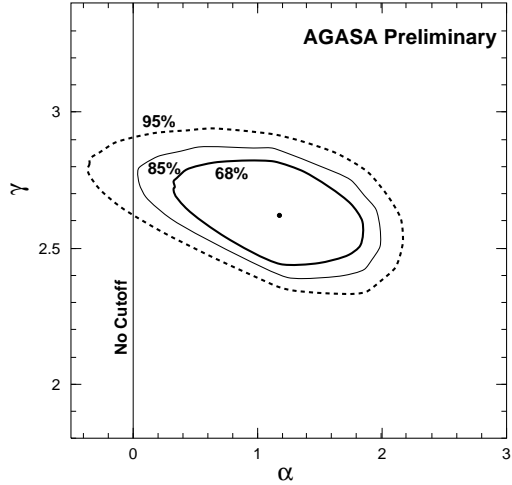}}}
\caption{The contour map showing the most likely values
of $\alpha$ and $\gamma$ and their significances estimated by
the likelihood analysis.}
\label{fig:contour_map}
\end{figure}

Figure \ref{fig:contour_map} shows the results of the likelihood
analysis. The most likely values of $\alpha$ and $\gamma$ are
1.18 and 2.62 respectively. The probability that the spectrum has
no cutoff ($\alpha\leq 0$) is calculated to be 14 \%
(In this case $\gamma$ is 2.76).  It is concluded that
the significance of a possible cutoff starting at the GZK cutoff
energy, $\epsilon_c = 4\times 10^{19}$ eV, is 85 \% C.L.  in the
present EHECR spectrum measured by AGASA.  This analysis has
indicated that the signature of a GZK cutoff may be present, but
the possibility that the spectrum has no cutoff cannot be ruled
out. The likelihood analysis shows that this case has a 15 \%
probability.

It should be noted that this likelihood analysis also estimated
the likely range of the spectrum power law index, $\gamma$, in
the presence of a cut-off (non-zero positive values of $\alpha$).
The 68 \% C.L. of the index is $2.62\pm 0.18$ which confirms the
harder spectrum of EHECRs compared to the spectrum at lower
energies \cite{yoshida95}.

\begin{figure}
\centerline{{\epsfxsize7.0cm\epsfbox{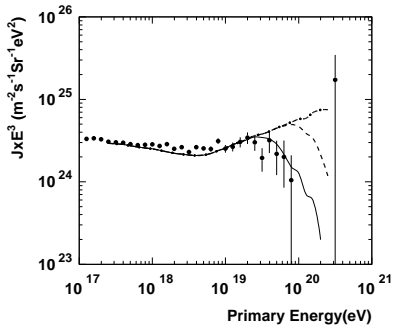}}}
\caption{ Fly's Eye monocular energy spectrum. Dots: data. Lines:
predicted spectra for source cutoffs at different energies. Solid
line: cutoff at $10^{19.6}$ eV. Dashed line: cutoff at
$10^{20}$eV. Chain line: cutoff at $10^{21}$ eV. }
\label{fig:fly_mono}
\end{figure}

In the Fly's Eye case, the stereo data set runs out of events
above $10^{19.4}$eV. Therefore, the Fly's Eye group must rely on
monocular data for the energy region above the ``ankle''.  The
monocular data set is much larger, but it relies on phototube
trigger times for event reconstruction.  The time-fitting tends
to yield larger geometrical errors leading to larger uncertainty
in energy determination. The group also made loose cuts on
the monocular spectrum, so as to keep the statistics as large as
possible and to avoid possible high energy event losses.  The
total energy spectrum is shown in Fig.\ref{fig:fly_mono}. Because
of the limited energy resolution, the differential energy
spectrum observed by the monocular eye (multiplied by $E^3$) does
not show the degree of structure found in the stereo data.
By using the two data sets (stereo and monocular),
the Fly's Eye group was able to check the cutoff without
selecting a spectral slope {\em a priori}, and at the same time
taking energy resolution into account.

   The following method was used. The group took the stereo
spectrum as the ``true'' energy spectrum because of its good
energy resolution, and predicted what the monocular Fly's Eye
would observe by using the actual monocular aperture and energy
resolution.  (The resolution was calculated by comparing stereo
and monocular estimates of energy for showers viewed in stereo.)
The curves shown in Fig.\ref{fig:fly_mono} are three expected
monocular spectra assuming a stereo spectrum cutoff at
$10^{19.6}$, $10^{20}$, and $10^{21}$eV respectively. It is easy
to see from the figure that the spectrum agrees well with a
cutoff at $10^{19.6}$ eV, with the exception of 
the highest energy event at $3.2\times10^{20}$eV.

Yakutsk also favors a cutoff in the spectrum. They expected 10
events above $10^{20}$ eV, but only one event was
recorded\cite{afan95a}. The highest energy event detected by
Yakutsk was estimated to have an energy of approximately
$1.5\times10^{20}$ eV\cite{egor93}. The event arrived with a very
large zenith angle (58.9$^o$) and traversed almost 2000$g/cm^2$
of atmosphere before reaching the detectors.  Therefore a large
attenuation correction to S(600) had to be applied, which leads
to the largest uncertainties in the energy estimation of this event.

In the case of Haverah Park, the group expected 8.3 events above
$4\times10^{19}$ eV ($10^{19.6}$ eV), and observed 15
events. Therefore, ``there is no evidence from these data for any
cut-off in the spectrum at energies above $4\times10^{19}$
eV''\cite{lawr91,wats91}.

    Table 2 lists the exposure and the number of events above
$10^{20}$ eV for each experiment. Among all the experiments,
Haverah Park has the highest number of events per exposure above
$10^{20}$ eV. We should remember here, however, that this ``traditional''
method of the cutoff search simply by 
using number of events above $10^{20}$ eV
has many problems as we discussed above. Furthermore,
comparison of the results between the different experiments would
require careful attentions concerning systematic uncertainty
in the energy scale in each experiment. We will mention this point
later.

\begin{table}
\begin{center}
\caption{Exposure at $10^{20}$eV and \# of events above $10^{20}$ eV }
\begin{tabular}{|c|c|c|c|}\hline 
Experiment  & exposure & \# of events above  &  \# of events significantly\\
 & ($km^2yrsr$) & $10^{20}$ eV &above $10^{20}$ eV\\
Haverah Park & 281 & 5 & 0 \\
AGASA & 790 & 2 & 1 \\
Yakutsk & 850 & 1 & 1 \\
Fly's Eye & 825 & 1 & 1 \\
\hline
\end{tabular}
\end{center}
\label{table:exposure}
\end{table}

Although both the Fly's Eye and AGASA measurements 
may favor the cutoff's existence, their detection
of super-energetic events well beyond the GZK energy has muddied
the simple GZK picture. We discuss these events next.

\section{ The Highest Energy Events}

    The detection of an event well beyond the GZK cutoff 
by the Fly's Eye group raised much interest. Shortly
after that, the AGASA group detected an event above 200 EeV. We
discuss these two events in this section.

\begin{figure}
\centerline{{\epsfxsize8cm\epsfbox{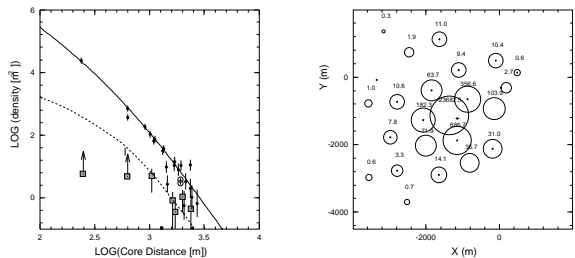}}}
\caption{Left:The lateral distribution of charged particles
(closed circles) and muons (squares) of the super energetic event
recorded by AGASA.  The large open circle is the density measured
by a detector designed to study the arrival time distribution of
particles in air showers.  Right: Map of the density distribution
of the event. A cross shows the estimated location of the shower
core.}
\label{fig:agasa_big}
\end{figure}

\begin{figure}
\centerline{{\epsfxsize7.0cm\epsfbox{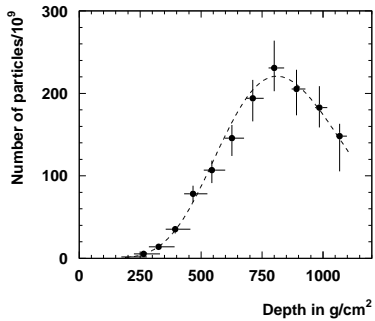}}}
\caption{ Shower longitudinal profile for the highest energy
event recorded by the Fly's Eye.}
\label{fig:fe320eev}
\end{figure}

AGASA would have demonstrated the GZK cutoff were it not for
the existence of its most energetic event, whose energy was estimated to be
$2.1^{+0.5}_{-0.4}\times 10^{20}$ eV \cite{hayashida94}.  The
details of the event are summarized in table 3.  This event hit
the array almost at its center, and 23 detectors surrounding the
shower core measured local electron densities and 
shower front arrival time.  Detectors more than 2 km away
from the core were triggered.  Thus, the reconstruction of the
lateral distribution of electrons for this event was excellent.
The functional fit of the electron density distribution agreed
well with all measured densities, which led to an error in
the estimation of the charged particle density at 600 meters
(S600), of +21 \% and -6.6 \%.  The largest uncertainty in the
energy estimation arises from the fact that the attenuation
length of S600 as a function of the atmospheric depth is not
well known.  The zenith angle of this event was $23^{\circ}$.  A
correction was made to convert the observed S600 to what
would be expected for a vertical shower.  Although the Monte
Carlo prediction suggests that the attenuation effect on S600 is
rather moderate, this effect cannot be measured because no other
events clearly above 100 EeV have been detected.  Furthermore, if this
event happened to be observed at the maximum of S600 development,
the density at 600 m may not be attenuated at all. Assuming no
attenuation gives a lower bound on the primary energy of 170
EeV. This energy is still well beyond the GZK cutoff.  

Five muon detectors recorded local muon densities within the
dynamic range of the counters. The estimated muon density at 600m
from the core is 42.8 m$^2$, which agrees well with $31\pm$ 9$
m^2$ obtained by extrapolation from the lower energy
region. There is no clear indication from the EAS muon content
that the primary particle is a gamma ray or an extremely heavy
nucleus.  All that can be claimed concerning this event is that
all features including the lateral distribution of electrons,
the muon density, and the timing distributions of the particles
are consistent with extrapolations from lower energy events.

\begin{table}[b]
\caption{Details of the most energetic events seen by AGASA and Fly's Eye }
\begin{center}
{\small
\begin{tabular}{|c|c|c|c|c|c|c|l|}\hline
    & Energy & R. A. & Dec. & Gal.Lat. & Gal.Long. & \
Zenith Angle & other parameters \\\hline

AGASA & $2.1\times 10^{20}$ eV& $18.9^{\circ}$ & $21.1^{\circ}$ & \
$-41^{\circ}$ & $131^{\circ}$ & $22.9^{\circ}$ & S(600)=892$m^{-2}$ \\
Fly's Eye & $3.2\times 10^{20}$ eV& $85.2^{\circ}$ & $48.0^{\circ}$ &\
$9.6^{\circ}$ & $163.4^{\circ}$ & $43.9^{\circ}$ & \
$X_{max}=815^{+60}_{-53}g/cm^2$ \\\hline
\end{tabular}
}
\end{center}
\label{table:big_events}
\end{table}

 On Oct. 15, 1991, The Fly's Eye observed an event \cite{loh93}
with an energy of $(3.2 ^{+0.92}_{-0.94}) \times 10^{20}$
eV\cite{bird95}.  This event impacted 13 kilometers away from
Fly's Eye I with a zenith angle of $43.9^o$ and an azimuth angle
of $31.7^o$. At the second site 3.4 km away, the partial eye
Fly's Eye II monitored that half of the visible sky
which was in the direction of Fly's
Eye I.  Unfortunately, this super high energy event landed on the
blind side of Fly's Eye II, so it was not observed stereoscopically.
Nonetheless, the event was particularly well measured by Fly's Eye I.
It fired 22 5.5$\times$5.5$^o$ photomultiplier tubes. The signals
were so strong that the high gain channels of several tubes were
saturated. The event also has a well-measured longitudinal shower
profile.  The $X_{max}$ of this event (Fig.\ref{fig:fe320eev}) is
estimated to be $815^{+60}_{-53} g/cm^2$, with most of the
uncertainty coming from the fit of the event geometry.  The best
estimate of $X_{max}$ falls between that expected for proton and
iron showers of this energy.  With a single event it is
particularly difficult to identify the particle as either
proton or iron. Indeed the Fly's Eye group could not rule
out the possibility of the event being a high energy $\gamma$.
No strong nearby source is obvious in its arrival direction:
right ascension $85.2^o \pm0.5^o$, declination $48.0^o$
$^{+5.2}_{-6.3}$\cite{bird95}. There are two candidate sources
near to the direction of the shower (3C147 \& MCG8-11-11), but
their distances appear too large for a 320 EeV event to
travel\cite{elbe95}. Rachen proposed that 3C134 might be a good
candidate but no redshift measurement is available
because of the galactic obscuration \cite{rachen94}.

If our understanding of the energy estimation on these events is
correct, we have another mystery.  How could these EHECRs reach
earth with such enormous energies?  It requires their sources to
be remarkably close, less than 50 Mpc, to prevent the expected
significant energy loss by the GZK mechanism.  Their super-high
energies allow the arrival directions to point back to the source
locations because the extragalactic magnetic field could not bend
their trajectories too much.  The upper bound on the deflection
angle of a $2\times 10^{20}$ eV event given by equation (4) is
only $\sim 1^{\circ}$ for a source distance of 50 Mpc, provided that 
larger inter-cluster magnetic fields do not actually exist
along the propagation path.  This is
comparable to, or smaller than, the angular resolutions of the
experiments. Many searches for possible astronomical sources have
been conducted \cite{elbe95}, but nothing interesting has been
found in the arrival directions of the highest energy
particles.

These events may, thus, suggest the appearance of a new
population of cosmic rays {\it above} the GZK cutoff.  This
possibility has triggered exotic ideas like the
topological defects (``top down'') model
\cite{bhattacharjee92,bhat95,sigl97} and the GRB model
\cite{waxman95} which we have already mentioned in this review.  The
top down model predicted arrivals at earth of gamma rays or
protons above the GZK cutoff, with energies originating from GUT
scale phenomena.  It is difficult to make any claims about the
primary composition of the highest energy events.  Halzen et
al.\cite{halz95} argued that the Fly's Eye event could not have
been initiated by a photon, based on its shower $X_{max}$. 
They claimed that a 3$\times 10^{20}$ eV photon initiates
an air shower whose peaks at $\sim 940 g/cm^2$ (without the LPM effect)
or $\sim 1075 g/cm^2$ (with the LPM effect).
It should be remarked, however, that the Fly's Eye shower was a
monocular event with an uncertainty in $X_{max}$ of about
60$g/cm^2$, mostly caused by uncertainty in geometry. 
In addition, the reconstruction of this event with deeper $X_{max}$
would lead to lower energy estimation, which decreases the differences 
between the observation and their Monte Carlo results.
Therefore, the Fly's Eye event by itself cannot rule out the photon origin.
On the same note, however, the AGASA event shows normal
muon distributions similar to events at lower energies. 
Thus, these showers are likely to be regular
hadronic showers.  
The GRB model predicts a correlation between
the energy and the arrival time, but the current statistics and
resolution do not allow a reliable study of this kind of
correlation.
We should note that the absolute
intensity of the flux above the GZK cutoff is consistent with
extrapolation of the flux below the cutoff.
We cannot give any natural reason why these models give the rate
of the super GZK events following the extrapolation.

These arguments concerning the ``new'' population of cosmic rays
are based mainly on these two 
events. Although the groups have carefully
analyzed the data, and nothing has been found to cause
unreasonable overestimation of energies in these events, we
should not neglect the possibility that an unpredicted tail of
the energy resolution may have given rise to these remarkable energies.
The energy estimation of the AGASA events relies mostly on Monte
Carlo studies and extrapolation of the observed behavior in the
lower energy region. The Fly's Eye event was a monocular one
(viewed only by a single eye).  Detection of more super-GZK events
with reliable energy estimation is required in order to say more
about this class of events. In addition, it is important to increase
number of events between $10^{19}$ and $10^{20}$ eV which are able
to ``calibrate'' our energy estimation for these highest energy events.

\section{ Absolute energy calibration and energy resolution}

   Before we leave the energy spectrum, we must discuss the absolute
energy calibration and energy resolution of the experiments.

   One question about the Fly's Eye energy determination is the
absolute efficiency for the production of 
fluorescence light.  All Fly's Eye analysis are based on
the efficiency estimate compiled by Bunner\cite{bunn67}.  A recent
measurement of the scintillation efficiency has been conducted by
Kakimoto et al\cite{kaki96}.  The Fly's Eye group have compared
the event energies using Bunner's efficiency and the new
efficiency, and have found the difference to be only about 1\%
\cite{loh96}.

    The insensitivity of the $\rho(600)$ to the interaction model
does not exempt its model dependence totally. The MOCCA (or
pre-MOCCA) models\cite{hill71} used by the Haverah Park
experiment, no doubt represented the best knowledge of hadronic
interactions in the 1970's, however recent Fly's Eye, AGASA and
Yakutsk data favor models which are more consistent with
a quicker dissipation of
energy (faster than the scaling models used by Haverah Park)
\cite{gais93,ding97,flet94}.  Evidence shows that the MOCCA
program, which was used to determine the relation between the
primary energy and $\rho(600)$, underestimates the muon density
by almost a factor of two at large distances from the core for
showers between 10$^{18}$ and 10$^{19}$ eV.  A water Cherenkov
detector is more sensitive to muons compared with very
soft electrons far from the shower core.  As discussed by
Lawrence, Reid, and Watson \cite{lawr91}, the overestimation in
energy could be as large as 40\% if, indeed, the high dissipation
model is correct.  The 40\% will certainly reduce the
disagreement between Haverah Park and the other experiments in
terms of the number of events above $10^{20}$eV per unit exposure
(see table 2). This would make the Haverah Park result more comparable
with a GZK cutoff.  We have seen arguments
that the systematics in energy calibration among experiments are
much smaller than 40\%, based on the flux differences.
However, the aperture calculation and
the energy resolution have a large effect on the estimated flux.
The flux can be significantly underestimated near the detector
threshold energy.  
This is because there are no events below
the threshold energy that trigger the array and are reconstructed
to have an energy higher than their true energy.  On the other
hand, showers just above the threshold can have reconstructed
energies above their true energy, and rob the threshold area of
flux due to the familiar ``downhill'' effect.

  We must also emphasize the
importance of resolution. The effect can easily be demonstrated
by observing the difference between the Fly's Eye monocular
spectrum (Fig.\ref{fig:fly_mono}) and the stereo spectrum
(Fig.\ref{fig:fly_stereo}).  Due to the poorer resolution, the
``dip'' structure in the monocular spectrum is much less striking
than that in the stereo spectrum.  The apparent ``dip'' position is also
shifted to nearly $10^{18.9}$ eV. The relative energy calibration
between the monocular and stereo events has been carefully
balanced using the stereo events, hence the shift in the 'dip'
position is entirely due to the effect of resolution. 

\section{ Chemical composition}

\subsection{ Composition from $X_{max}$ measurements}

    The position of shower maximum in the atmosphere ($X_{max}$)
in $g/cm^2$ is sensitive to the composition of the primary
particle. Protons for instance, will on average experience their
first interaction deeper in the atmosphere than heavy nuclei of
the same energy.  Proton showers are also expected to develop
more slowly than heavy primary showers with the same energy per
primary particle. The primary chemical composition can
be, therefore, deduced from the distribution of $X_{max}$.

    The Fly's Eye group derive their composition estimate by
comparing the measurements with Monte Carlo predictions. The
Monte Carlo showers are generated using a QCD Pomeron model ( the
so-called KNP model) \cite{gais93,kope89}. The Monte Carlo
generated showers are then passed through the detector Monte
Carlo simulation program to account for detector trigger biases. Those
events triggering both Fly's Eye sites in the detector Monte
Carlo are written to a data file with the same format as the real
data.  This fake data set is then reconstructed using the same
programs used in the real data analysis.

\begin{figure}
\centerline{{\epsfxsize7cm\epsfbox{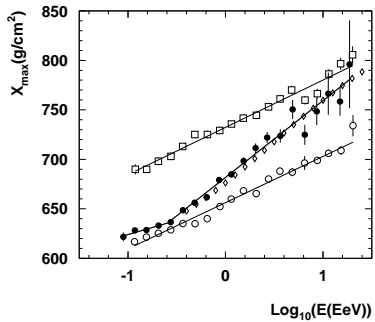}}}
\caption{ The $X_{max}$ elongation rate. Black dots: Fly's Eye
data.  Open squares : proton $X_{max}$ distribution based on the
KNP model.  Open circles: iron $X_{max}$ distribution based on
the KNP model.  Diamonds: expected mean $X_{max}$ distribution
based on a simple two component model.}
\label{fig:fly_xm}
\end{figure}
 
The mean X$_{max}$ as a function of primary energy measured by
the Fly's Eye detectors, is shown in Fig.\ref{fig:fly_xm}
together with KNP model Monte Carlo generated proton and iron showers.
From the figure, one can see that the
composition is heavy at a few times 10$^{17}$ eV and gradually
shifts to light primaries near 10$^{19}$eV.  The same conclusion
is reached by comparing the rise and fall of the full X$_{max}$
distributions in each energy bin \cite{soko93}.  The elongation
rate (the increment of $X_{max}$ per decade of energy) from 0.3
EeV to 10 EeV is 78.9$\pm$3$g/cm^2$ per decade for the real data,
and 50 $g/cm^2$ per decade for the Monte Carlo simulated proton
or iron showers.

Constraints on hadronic interaction models by the Fly's Eye
measurements arise from the fact that the Fly's Eye measures both
the absolute $X_{max}$ position at each energy and the elongation
rate. The absolute mean value of $X_{max}$ around
$3\times10^{17}$ eV (about 630$g/cm^2$) essentially rejects any
model with a large elongation rate, since those large elongation
rate models inevitably predict a deeper $X_{max}$ at
$3\times10^{17}$ eV, even with an iron primary.

   The facts that the measured absolute value of $X_{max}$ at
$3\times10^{17}$ eV is low and that the measured elongation rate
is high, naturally leads to the conclusion that the composition
is becoming lighter over the energy range observed.  Of course, a
quantitative prediction of how quickly the composition gets
lighter is still model dependent.

A recent result from the HiRes prototype detector supports the
conclusion that the composition around $3\times10^{17}$ eV is
heavy\cite{kidd97}.  In addition, event reconstruction using the
new air scintillation efficiency (mentioned earlier) does not
affect the original Fly's Eye composition conclusion.

\begin{figure}
\centerline{{\epsfxsize7cm\epsfbox{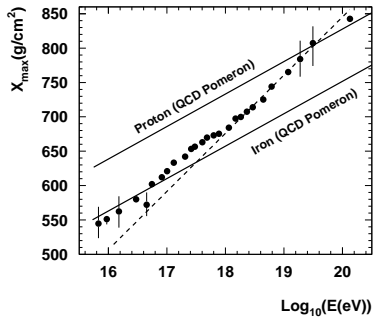}}}
\caption{ The Yakutsk mean $X_{max}$ as a function of primary
energy.  Solid lines are the Monte Carlo predictions of a QCD
pomeron model from the Fly's Eye group. The dashed line is the fit to
points above $1.1\times10^{18}$ eV.}
\label{fig:yakutsk_xm}
\end{figure}

  Yakutsk derives the depth of shower maximum using the Cherenkov
lateral distribution\cite{dyak93,egor93}.  Parameters used to get
$X_{max}$ are the ratio of the Cherenkov flux to the charged
particle size, the Cherenkov lateral distribution slope at core
distances between 100 and 400 m, and a characteristic radius,
$R_o$, determined by the Cherenkov light between 50 and 300
m from the core \cite{dyak93}.  
Further communication from the Yakutsk group is
still necessary to fully understand how these parameters are used
to derive $X_{max}$. Fig.\ref{fig:yakutsk_xm} shows the derived
$X_{max}$ as a function of primary energy\cite{egor93}. We
realized at the time of writing this paper, that the Monte Carlo
predictions in the Yakutsk $X_{max}$ plot are digitized from the
Fly's Eye $X_{max}$ plot.  Around $10^{17}$ eV, the Yakutsk group
claim their $X_{max}$ values indicate that the composition is
mixed. They further claim that 
as energy increases the composition gets lighter. The QCD
Pomeron elongation rate is about 50$g/cm^2$ per decade over this
energy range. The measured elongation rate is $79\pm3g/cm^2$ per
decade for events above 1.1$\times10^{18}$ eV\cite{egor93}.  But
we need to be very careful here since the Monte Carlo predictions
include the Fly's Eye detector bias and the Yakutsk Cherenkov
detector does not necessarily have the same detector bias as the
Fly's Eye fluorescence detector.  We do notice that a qualitative
conclusion which agrees with the above picture has been drawn by the
same group\cite{afan96}.  In this case, the conclusion is based
on the QGS\cite{kalm95} model, but no plot is shown. The typical
separation in mean $X_{max}$ between proton and iron showers is
about 75 to 100 $g/cm^2$ in this energy range, and the detector
bias could be as large as 20 or 30$g/cm^2$. Therefore, we
encourage the Yakutsk group to carry out their own detector Monte
Carlo simulations when comparing their data with predictions.

\subsection{ Composition from muon to electron ratio}

The AGASA experiment has measured the muon density as a function
of the primary energy of cosmic rays or rather as function of S600,
their observable energy estimator. Their results at higher energies are
very consistent with that expected from 
extrapolation of data in the lower energy
region. They measured the slope of the logarithm of muon density
to that of electrons and found that this slope does not change
for E$\geq 10^{17.5}$ eV.  They used the MOCCA event generator
package \cite{hill71} to find that the Fly's Eye's picture should
have caused a change in the slope \cite{hayashida95}.  It is
claimed that there is
no evidence in the AGASA measurement to support the Fly's Eye
result of a change in the chemical composition. 

\begin{figure}
\centerline{{\epsfxsize6.5cm\epsfbox{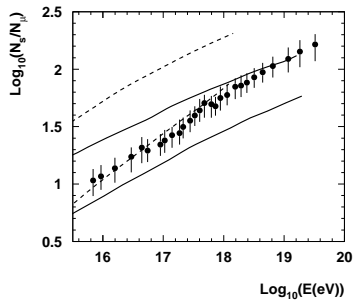}}}
\caption{ The ratio of charged particle size to muon size (muon
threshold 1 GeV) as a function of the primary energy from the
Yakutsk experiment.  Solid lines are the Monte Carlo predictions
of the QGS model (Upper curve: proton, lower curve: iron). Dashed
lines are the Monte Carlo predictions of a scaling model(Upper
curve: proton, lower curve: iron).}
\label{fig:yakutsk_mu}
\end{figure}

    In addition to the Cherenkov lateral distribution, the
Yakutsk group also measure muon densities. Instead of plotting
muon size against the all charged particle size as is usually
seen, the reverse is plotted in Fig.\ref{fig:yakutsk_mu}.  Two
model predictions are plotted in the same figure, one uses the
QGS model \cite{kalm95}, and the other uses an old-fashioned
scaling model\cite{elbe79}.  It is interesting to note that the
QGS model predicts a charged particle to muon ratio which is more
than a factor of two smaller than the scaling model prediction at
energies above $10^{18}$ eV.  The Yakutsk group reach the same conclusion
as the Fly's Eye 
about the composition as their $X_{max}$ result, if the QGS model
is used to interpret the data.  On the other hand, the scaling
model predicts fewer muons, and prefers a flux of almost 100\%
iron below $10^{18}$ eV before running into difficulty above
$10^{18}$eV where a composition heavier than iron is required.
Here again we emphasize the importance of a detector Monte
Carlo. It is vital to take the detector bias into account when
comparing with Monte Carlo predictions.

\subsection{ Effort in unifying the composition results}

\begin{figure}
\centerline{{\epsfxsize8.0cm\epsfbox{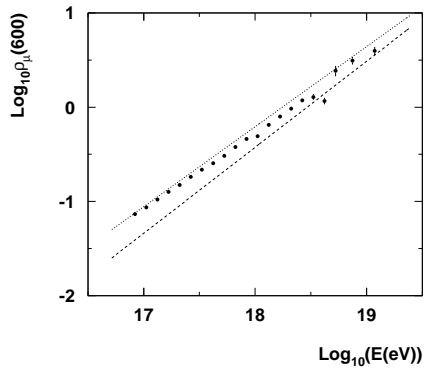}}}
\caption{ The muon density at 600 meters from shower core as a
function of the primary energy from the Akeno A1 array. The lines
are the predictions of 'Sibyllized MOCCA'.  Dotted line: iron
primaries. Dashed line: proton.}
\label{fig:brd_mu}
\end{figure}

The apparent contradiction between the Fly's Eye $X_{max}$ measurement
and the AGASA muon to electron ratio measurement draws much
attention.  The interaction model (contained in the original
MOCCA) used by AGASA to interpret the muon data was immediately
questioned.  From the absolute $X_{max}$ position at low energy
(3$\times10^{17}$eV), the Fly's Eye group and their model
providers \cite{gais93,ding97} realized that no scaling model
fits the data at this energy.  A model with a higher rate
of energy dissipation, either through large cross-section or
through large inelasticity, or both, is required. Such a fast
energy dissipation model leads to the rapid development of
showers, smaller depth of shower maximum ($X_{max}$) and a
smaller elongation rate (the increment of $X_{max}$ per decade
of energy). Hence, in terms of shower development, a proton shower
will asymptotically approach a ``conventional'' iron shower,
where ``conventional'' means an extrapolation from lower energy
behavior using a scaling model.  Similarly, in terms of muon
content proton showers will become more and more like
conventional iron showers, since both $X_{max}$ and the muon
content are related to the distributions of secondary particle
energies.

This inspired a group of people from Adelaide (Dawson et. al.
\cite{daw98}) to
perform a cross check using the so called ``Sibyllized MOCCA'',
which uses the MOCCA program as the shower driver, but uses the
Sibyll model\cite{flet94} for the hadronic interactions. Their
simulations show that while exact fractions of protons and iron derived
from the Fly's Eye measurements of $X_{max}$ are somewhat model
dependent, there is still clear evidence that the composition is
changing from heavy to light.  The elongation rate from this
model is smaller than that predicted by the original MOCCA code
(without Sibyll).  This has a related effect for muons: the newer
model gives a steeper slope to the muon content as a function of
energy. In fact the proton slope is even steeper than that for
iron at the same energy per primary particle. The energy per
nucleon for protons is 56 times that of iron. With the Sibyll
model the AGASA muon measurement no longer contradicts a changing
composition (see Fig.\ref{fig:brd_mu}).

As a by-product of composition measurements, it is now believed
that the cascade of nucleon-nuclei interactions dissipates energy
faster than the scaling models would predict using an
extrapolation from lower energies.

To summarize the composition measurements, we have seen that the
Fly's Eye $X_{max}$ measurements indicate that the composition is
shifting from heavy to light over the energy range from $10^{17}$
eV to $10^{19}$ eV. At Yakutsk, both the $X_{max}$ and the muon
density results favor a composition change from a mixture of
heavy and light to light over the same energy region, however, a
detector Monte Carlo is needed to strengthen their conclusions.

While the original analysis of the AGASA muon measurement states
that there is no indication of a changing composition, the application
of a consistent hadronic model brings the results into better
agreement with the Fly's Eye conclusion. 
In terms of the composition, the Fly's Eye results
support a two component picture where a heavy flux is
progressively dominated by a protonic flux at higher energies.
Yakutsk does not support the first component being very heavy,
but does favor a light second component.  AGASA cannot be
regarded as supportive of a two component picture, based on their
original analysis, but there may be room for a consensus if
conclusions are based on similar hadronic models.
It is clear, however, that better measurements and a further refinement
of interaction models are necessary to resolve the composition
issue.

\section{ Anisotropy}

\begin{figure}
\centerline{{\epsfxsize8cm\epsfbox{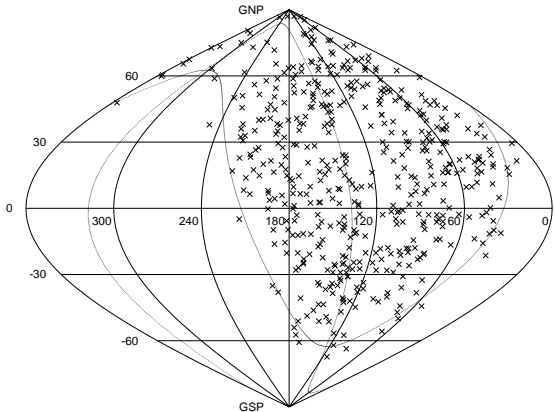}}}
\caption{Arrival directions (in galactic coordinates) of EHECRs
above $10^{19}$ eV detected by AGASA.}
\label{fig:agasa_direction}
\end{figure}

Results from all experiments indicate that cosmic ray arrival
directions are largely isotropic.  Typical upper bounds on the
amplitude of the anisotropy (first and second harmonics in right
ascension) are less than 5\% for events around $10^{18}$ eV, less
than 10\% for events around $10^{18.5}$ eV and less than 30\% for
events around $10^{19}$eV\cite{tesh93}.  Lee and Clay\cite{lee95}
have argued that, based on the amplitudes of the
harmonics, cosmic rays in this energy range cannot be protons
originating within the galaxy.

Based on a data set mainly from Haverah Park, Stanev et
al\cite{stan95} pointed out that the arrival directions of cosmic
rays with energies above $4\times10^{19}$eV (a total of 42
events) exhibit a correlation with the general direction of the
supergalactic plane, a plane defined by nearby radio galaxies
($z\leq 0.02$) in the northern hemisphere. The chance probability
that a uniform distribution would have such a concentration is a
few percent, according to their simulation.  This concentration
diminishes as the cosmic ray energy goes down.  There is no
galactic plane enhancement (the same analysis actually indicates
that cosmic rays above $4\times10^{19}$ eV are more likely to
come from large galactic latitudes).  Kewley, Clay, and
Dawson\cite{kewl96} applied the same analysis to the southern
part of the sky using data from the Sydney SUGAR array
\cite{winn86} and no concentration around the supergalactic
plane was found.  This may be because the plane is less well
defined in the south.

\begin{figure}
\centerline{{\epsfxsize7cm\epsfbox{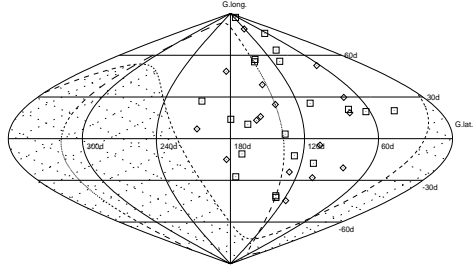}}}
\caption{Arrival directions in galactic coordinate of EHECRs
above $4\times 10^{19}$ eV recorded by AGASA (Open squares --
$E\ge 5\times 10^{19}$ eV, open circles -- $5\times 10^{19}$ eV
$\ge E \ge 4\times 10^{19}$ eV).
The dashed curve shows the supergalactic
plane. The cross hatched area represents parts of the sky
unobservable by AGASA.}
\label{fig:agasa_pair}
\end{figure}

Figure \ref{fig:agasa_direction} shows the arrival directions of
all EHECRs recorded by AGASA with energies above $10^{19}$ eV. 
The distribution
is completely consistent with an isotropic distribution
\cite{hayashida97b}.  No evidence of enhancement associated with
the galactic plane has been detected and the group did not find
significant enhancement along the supergalactic plane
\cite{hayashida96}.  However, an interesting feature has been
reported \cite{hayashida96}: event clusters.  Figure
\ref{fig:agasa_pair} shows the arrival directions of EHECRs above
$4\times 10^{19}$ eV recorded by AGASA.  Three event pairs each with
angular separations of less than $2.5^{\circ}$
have been detected.  Two of the pairs are within $2^{\circ}$ of
the supergalactic plane.  The chance probability of having
two or more such pairs has been
estimated to be $2.9\times 10^{-2}$.

None of these pairs arrive from our galactic plane again favoring
the extragalactic origins of EHECRs. However, these features
themselves contain some mysteries.  First, the AGASA group have
found no active astrophysical objects in the directions of the
pairs.  The threshold of $4\times 10^{19}$ eV requires sources
within $\sim 500$ Mpc \cite{yosh93} because of the energy loss in
the microwave background field. Some
Markarian-type galaxies have been found, but no FR II type radio
galaxies.  Second, if these events are protons, then possible
extragalactic magnetic fields should bend their trajectories,
which would result in a larger separation angle between the two
events in a pair.  Equation (4) gives 6.8$^{\circ}$ for sources
at a distance of 80 Mpc (corresponding to radio galaxies in the
supergalactic plane) and 17$^{\circ}$ for those at 500 Mpc (the
upper bound on the distance of possible sources).  To explain the
observed angular separation, then either the extragalactic
magnetic field must be much lower than the current upper bound given
in equation (3), or the primary particles in these events must be
neutrally charged, perhaps photons. The observed muon
densities in these events are, however, 
consistent with hadron primaries.

We should note here that this kind of analysis might be
criticized in terms of the uncountable degrees of freedom in the
estimation of the statistical significance. There is no clear
reason why an angular separation of 2.5$^{\circ}$ should be
chosen to define an event pair. It has been claimed that this
number is consistent with the experimental angular resolution,
but $2^{\circ}$ or 3.5$^{\circ}$ instead of 2.5$^{\circ}$ would
be just as consistent, within uncertainties, in the angular
resolution, which vary from event to event.  Furthermore,
a search for event clusters with a larger angular separation
might also be reasonable given the possible bending of
trajectories by the extragalactic field.  If the angular
separation can be considered as a free parameter, the chance
probability of 2.9\% might be too weak to claim the existence of
the pairs.  The same attention should be paid to the combined
analysis of the different experiments by Uchihori et.al.
\cite{uchi96}. They claimed a possible correlation of the event
clusters (including triplets) and the supergalactic plane with a
significance level of a few percent. It is true that the pooling
of results from different experiments is a vital effort aimed at
overcoming the limitation of poor statistics, but the problem of
dealing with data sets with different angular resolutions makes
it difficult to draw a reliable conclusion.  In contrast to the
angular separation, the choice of the threshold energy of
$4\times 10^{19}$ eV in the analysis can be justified since
this energy is the universal value marking the
beginning of the GZK cutoff in the spectrum.

\section{Summary}

  The EHE cosmic ray energy spectrum steepens in the energy
region between $10^{17.6}$ and $10^{18}$eV (the second ``knee'',
where the spectral slope changes from -3.0 to -3.3) and flattens
between $10^{18.5}$ and $10^{19}$ eV (the so called ``ankle'',
where the spectral slope changes from -3.3 to approximately
-2.7). The straightforward, and less model-dependent,
interpretation is a two component scenario: a high energy
extragalactic component dominates over a steeper galactic
component above the ankle.  The many experimental results now
available are supportive of this picture, including: a possible 
signature of the GZK cutoff obtained by the
Fly's Eye, AGASA and Yakutsk; an indication of the chemical
composition getting lighter at high energies from the Fly's Eye
and Yakutsk groups; no enhancement of the arrival direction
distribution associated with the galactic plane; a possible
correlation with the supergalactic plane found in a combined data
set mainly consisting of Haverah Park data; and three event
clusters above $4\times 10^{19}$ eV observed by AGASA arriving
from directions well away from the galactic plane.  

Although our two component picture seems to make sense
when we put these results {\it together}, the conclusion is far
from solid.  The significance of the GZK cutoff is muddied by the
super-high energy events well beyond the cutoff, thereby providing
complications to the simple picture of the GZK mechanism.  The
interpretation of chemical composition measurements has some
model dependence as cautioned by the AGASA results.  The
statistical significance of the event clusters only allows us to
suggest possible ``hints'' of something exciting.  What
encourages us about the two component scenario is the fact that
different analysis from different experiments seem to be
reasonably consistent under our scenario.
The next step is to make all these results robust by
accumulating more data with good resolution. For example, a
fine measurement of the GZK cutoff would clarify the
extragalactic hypothesis. A clear measurement of the mass
composition above the ``ankle'' would also be very helpful in
confirming or rejecting our current picture. A detection of a EHE
photon or neutrino component would bring us a new understanding
of the universe.  During the next decade, we will see the study of
EHE cosmic rays continue to provide a laboratory for
non-accelerator particle physics, and we look forward to it
establishing a new astronomy.

\vspace{0.5cm}

\centerline{\Large{\bf Acknowledgments}}

The authors are grateful to B. Dawson, P. Sommers, P. Sokolsky,
M. Teshima, J. N. Matthews, and G. B. Yodh for helpful suggestions
and advice. They also wish to thank M. Nagano
for allowing us to use the preliminary analysis by one of
the authors (S.Y.) based on the published data
from the AGASA experiment.


\end{document}